\documentclass[11pt]{article}
\usepackage{epsfig}
\usepackage{here}
\usepackage{a4}
\usepackage{amssymb}

\newcommand{\ANKARA}      {1}
\newcommand{\ANNECY}      {2}
\newcommand{\ASSERGILNGS} {3} 
\newcommand{\BARI}        {4}
\newcommand{\BERN}        {5}
\newcommand{\BOLOGNA}     {6}
\newcommand{\BRUSSELS}    {7}
\newcommand{\DUBNA}       {8}
\newcommand{\FRASCATI}    {9}
\newcommand{\FUNABASHI}  {10}	
\newcommand{\HAIFA}      {11}
\newcommand{\HAMBURG}    {12}
\newcommand{\GAZWADONG}  {13}
\newcommand{\KARIYA}     {14} 
\newcommand{\KOBE}       {15}
\newcommand{\LAQUILA}    {16}
\newcommand{\LYON}       {17}
\newcommand{\MOSCOWINR}  {18}
\newcommand{\MOSCOWITEP} {19}
\newcommand{\MOSCOWLPI}  {20}
\newcommand{\MOSCOWSINP} {21}
\newcommand{\MUNSTER}    {22}
\newcommand{\NAGOYA}     {23}
\newcommand{\NAPOLI}     {24}
\newcommand{\NEUCHATEL}  {25}
\newcommand{\OBNINSK}    {26}
\newcommand{\ORSAY}      {27}
\newcommand{\PADOVA}     {28}
\newcommand{\ROMA}       {29}
\newcommand{\ROSTOCK}    {30}
\newcommand{\SALERNO}    {31}
\newcommand{\SOFIA}      {32}
\newcommand{\STRASBOURG} {33}
\newcommand{\TUNIS}      {34}
\newcommand{\UTSUNOMIYA} {35}
\newcommand{\ZAGREB}     {36}
\newcommand{\ZURICH}     {37}

\newcommand{\OperaInstitutes}{
\ANKARA      . METU-Middle East Technical University, TR-06531 Ankara, Turkey \\
\ANNECY      . LAPP, Universit\'e de Savoie, CNRS/IN2P3, 74941 Annecy-le-Vieux, France\\
\ASSERGILNGS . Laboratori Nazionali del Gran Sasso dell'INFN, 67010 Assergi (L'Aquila), Italy \\
\BARI        . Dipartimento di Fisica dell'Universit\`a  di Bari and INFN, 70126 Bari, Italy \\
\BERN        . University of Bern, CH-3012 Bern, Switzerland \\
\BOLOGNA     . Dipartimento di Fisica dell'Universit\`a  di Bologna and INFN, 40127 Bologna, Italy \\
\BRUSSELS    . IIHE-Inter-University Institute for High Energies, Universit\'e Libre de Bruxelles, B-1050 Brussels, Belgium \\
\DUBNA       . JINR-Joint Institute for Nuclear Research, 141980 Dubna, Russia \\
\FRASCATI    . Laboratori Nazionali di Frascati dell'INFN, 00044 Frascati (Roma), Italy \\
\FUNABASHI   . Toho University, 274-8510 Funabashi, Japan \\
\HAIFA       . Department of Physics, Technion, 32000 Haifa, Israel\\ 
\HAMBURG     . Hamburg University, 22043 Hamburg, Germany\\
\GAZWADONG   . Gyeongsang National University, 900 Gazwa-dong, Jinju 660-300, Korea\\
\KARIYA      . Aichi University of Education, 448 Kariya (Aichi-Ken), Japan\\
\KOBE        . Kobe University, 657 Kobe, Japan \\
\LAQUILA     . Dipartimento di Fisica dell'Universit\`a  dell'Aquila and INFN, 67100 L'Aquila, Italy\\
\LYON         . IPNL, Universit\'e Claude Bernard Lyon 1, CNRS/IN2P3, 69622 Villeurbanne, France\\
\MOSCOWINR   . INR-Institute for Nuclear Research of the Russian Academy of Sciences, 117312 Moscow, Russia\\
\MOSCOWITEP  . ITEP-Institute for Theoretical and Experimental Physics, 117259 Moscow, Russia \\
\MOSCOWLPI   . LPI-Lebedev Physical Institute of the Russian Academy of Sciences, 117924 Moscow, Russia\\
\MOSCOWSINP  . SINP MSU-Skobeltsyn Institute of Nuclear Physics of Moscow State University, 119992 Moscow, Russia \\
\MUNSTER     . University of M\"unster, 48149 M\"unster, Germany\\
\NAGOYA      . Nagoya University, 464-01 Nagoya, Japan\\
\NAPOLI      . Dipartimento di Fisica dell'Universit\`a Federico II di Napoli and INFN, 80125 Napoli, Italy \\
\NEUCHATEL   . Universit\'e de Neuch\^atel, CH 2000 Neuch\^atel, Switzerland\\
\OBNINSK     . Obninsk State University, Institute of Nuclear Power Engineering, 249020 Obninsk, Russia\\
\ORSAY       . LAL, Universit\'e Paris-Sud 11, CNRS/IN2P3, 91898 Orsay, France\\
\PADOVA      . Dipartimento di Fisica dell'Universit\`a  di Padova and INFN, 35131 Padova, Italy \\
\ROMA        . Dipartimento di Fisica dell'Universit\`a  di Roma ``La Sapienza" and INFN, 00185 Roma, Italy \\
\ROSTOCK     . Fachbereich Physik der Universit\"at Rostock, 18051 Rostock, Germany\\
\SALERNO     . Dipartimento di Fisica dell'Universit\`a  di Salerno and INFN, 84084 Fisciano, Salerno, Italy \\
\SOFIA       . Faculty of Physics, Sofia University ``St. Kliment Ohridski", 1000 Sofia, Bulgaria\\
\STRASBOURG   . IPHC, Universit\'e Louis Pasteur, CNRS/IN2P3, 67037 Strasbourg, France\\
\TUNIS       . UPNHE-Unit\'e de de Physique Nucl\'eaire et des Hautes Energies, 1060 Tunis, Tunisia \\
\UTSUNOMIYA  . Utsunomiya University, 320 Tochigi-Ken, Utsunomiya, Japan\\
\ZAGREB      . IRB-Rudjer Boskovic Institute, 10002 Zagreb, Croatia\\
\ZURICH      . ETH-Eidgen\"ossische Technische Hochschulen Z\"urich, CH-8092 Zurich, Switzerland \\
}

\newcommand{\OperaAuthorList}{
R.~Acquafredda$^{\NAPOLI}$,
N.~Agafonova$^{\MOSCOWINR}$,
M.~Ambrosio$^{\NAPOLI}$,
A.~Anokhina$^{\MOSCOWSINP}$,
S.~Aoki$^{\KOBE}$,
A.~Ariga$^{\NAGOYA}$,
L.~Arrabito$^{\LYON}$,
D.~Autiero$^{\LYON}$,
A.~Badertscher$^{\ZURICH}$,
A.~Bergnoli$^{\PADOVA}$
% ...
\\
}

\begin{document}

\def\numunue{\nu_\mu\rightarrow\nu_e}
\def\numunutau{\nu_\mu\rightarrow\nu_\tau}
\def\nuebar{\bar\nu_e}
\def\nue{\nu_e}
\def\nutau{\nu_\tau}
\def\numubar{\bar\nu_\mu}
\def\numu{\nu_\mu}
\def\ra{\rightarrow}
\def\numubarnuebar{\bar\nu_\mu\rightarrow\bar\nu_e}
\def\nuebarnumubar{\bar\nu_e\rightarrow\bar\nu_\mu}
\def\osc{\rightsquigarrow}

\def\inteni{{\cal I}_{pot}}
\def\fmerit{{\cal F}}

\title{\bf First events from the CNGS neutrino beam detected\\ in the OPERA experiment}

\maketitle

\author{\noindent \\ \OperaAuthorList }

\begin{flushleft}
\footnotesize{\OperaInstitutes }
\end{flushleft}

\baselineskip=14pt

\vspace{0.2cm}

\begin{abstract}

The OPERA neutrino detector at the underground Gran Sasso Laboratory (LNGS) was designed 
to perform the first detection of neutrino oscillations in appearance mode, through the study
of  $\numunutau$ oscillations.
The apparatus consists of a lead/emulsion-film target complemented by electronic detectors.
It is placed in the high-energy, long-baseline CERN to LNGS beam (CNGS) 730 km away from the 
neutrino source.
In August 2006 a first run with CNGS neutrinos was successfully conducted. A first sample of neutrino events was collected, 
statistically consistent with the integrated beam intensity. After a brief description
of the beam and of the various sub-detectors, we report on the achievement
of this milestone, presenting the first data and some analysis results.

\end{abstract}

\section{Introduction}
\indent

The solution of the long-standing solar and atmospheric neutrino puzzles has come from the hypothesis of
neutrino oscillations. This implies that neutrinos have non vanishing and not degenerate masses, 
and that their flavor eigenstates involved in weak interaction processes are a superposition of their mass eigenstates~\cite{osc}. 

Several key experiments conducted in the last decades with solar neutrinos
(see~\cite{bahcall} for a review), and with atmospheric, reactor and accelerator neutrinos, have contributed to build-up our present understanding of neutrino mixing. 
Atmospheric neutrino oscillations, in particular, have been studied by the Super-Kamiokande~\cite{sk}, 
Kamiokande~\cite{kam}, MACRO~\cite{macro} and SOUDAN2~\cite{soudan2} experiments. 
Long baseline experiments confirmed the oscillation hypothesis with accelerator neutrinos: K2K~\cite{k2k} 
in Japan and MINOS~\cite{minos} in the USA. 
The CHOOZ~\cite{chooz} and Palo Verde~\cite{paloverde} reactor experiments excluded 
the $\numunue$ channel as the dominant one in the atmospheric sector.

However, the direct appearance of a different neutrino flavor is still an important open issue. 
Long-baseline accelerator neutrino beams can be used to probe the atmospheric neutrino signal 
and confirm the preferred solution of 
$\numunutau$ oscillations. In this case, the beam energy should be large enough to produce the heavy $\tau$ lepton.
This is one of the main goals of the OPERA experiment~\cite{opera} that uses the long baseline (L=730 km)
CNGS neutrino beam~\cite{cngs} from CERN to LNGS, the largest underground 
physics laboratory in the world. 
The challenge of the experiment is to measure the appearance of $\nutau$ from $\numu$ oscillations.
This requires the detection of the short-lived $\tau$ lepton (c$\tau$ = 87.11 $\mu$m) with high efficiency and low background.
The $\tau$ is identified by the detection of its characteristic decay topologies, in one prong
(electron, muon or hadron) or in three-prongs.
The $\tau$ track is measured with a large-mass sampling-calorimeter made of 1 mm thick lead plates 
(absorber material) inter-spaced with thin emulsion films (high-accuracy tracking devices). 
This detector is historically called 
Emulsion Cloud Chamber (ECC)~\cite{opera}. Among past applications it was successfully used in the DONUT experiment 
for the first direct observation of the $\nutau$~\cite{donut}.

The OPERA detector is made of two identical Super Modules each consisting of a target section 
of about 900 ton made of lead/emulsion-film ECC modules (bricks), of a scintillator tracker detector, 
needed to pre-localize neutrino interactions within the target, and of a muon spectrometer.

The construction of the CNGS beam has been recently completed and a first run took place in August 2006
with good performance of the facility. 
First data were collected by the OPERA detector still without ECC bricks 
installed, yielding a preliminary measurement of the 
beam features along with the collection of a number of neutrino interactions (319) consistent with 
the integrated beam intensity of 7.6 $\times$ 10$^{17}$ protons on target (p.o.t.).
The OPERA experiment operated very satisfactorily during the run.


\section {The CNGS beam and the OPERA experiment}
\indent

The CNGS neutrino beam was designed and optimized for the study of $\numunutau$ oscillations in appearance mode,
by maximizing the number of charged current (CC) $\nutau$  interactions at the LNGS site. 
A 400 GeV proton beam is extracted from the CERN SPS in 10.5 $\mu$s short pulses with design intensity of 
2.4 $\times$ 10$^{13}$ p.o.t. per pulse. 
The proton beam is transported through the transfer line TT41 to the CNGS target T40~\cite{cngs}. 
The target consists of a series of thin graphite rods helium-cooled. 
Secondary pions and kaons of positive charge produced in the target are focused into a parallel beam by a system of two 
magnetic lenses, called horn and reflector.  A 1,000 m long decay-pipe allows the pions and kaons to decay 
into muon-neutrinos and muons. The remaining hadrons (protons, pions, kaons) are absorbed by an iron beam-dump. 
The muons are monitored by two sets of detectors downstream of the dump;  they 
measure the muon intensity, the beam profile and its center. 
Further downstream the muons are absorbed in the rock while neutrinos continue their travel towards Gran Sasso.

The average neutrino energy at the LNGS location is $\sim$ 17 GeV.
The $\numubar$ contamination is $\sim$ 4\%, the  $\nue$ and $\nuebar$ contaminations are lower than 1\%,
while the number of prompt $\nutau$ from D$_s$ decay is negligible. The average L/E$_{\nu}$  ratio is 43 km/GeV. 
Due to the earth curvature neutrinos from CERN enter the LNGS halls with an angle of about 3$^\circ$
with respect to the horizontal plane.

Assuming a CNGS beam intensity of 4.5 $\times$ 10$^{19}$ p.o.t. per year and a five year 
run about 31,000 CC plus neutral current (NC) neutrino events will be collected by OPERA
from interactions in the lead-emulsion target. 
Out of them 95 (214) CC $\nutau$ interactions are expected for oscillation parameter values
 $\Delta m^2_{23}$= 2 $\times$ 10$^{-3}$ eV$^2$ (3 $\times$ 10$^{-3}$ eV$^2$) and sin$^2$2$\theta_{23}$=1. 
Taking into account the overall $\tau$ detection efficiency the experiment should gather 
10-15 signal events with a background of less than one event.

In the following, we give a brief description of the main 
components of the OPERA detector. 

\begin{figure}[ht]
\begin{center}
\includegraphics[width=13cm]{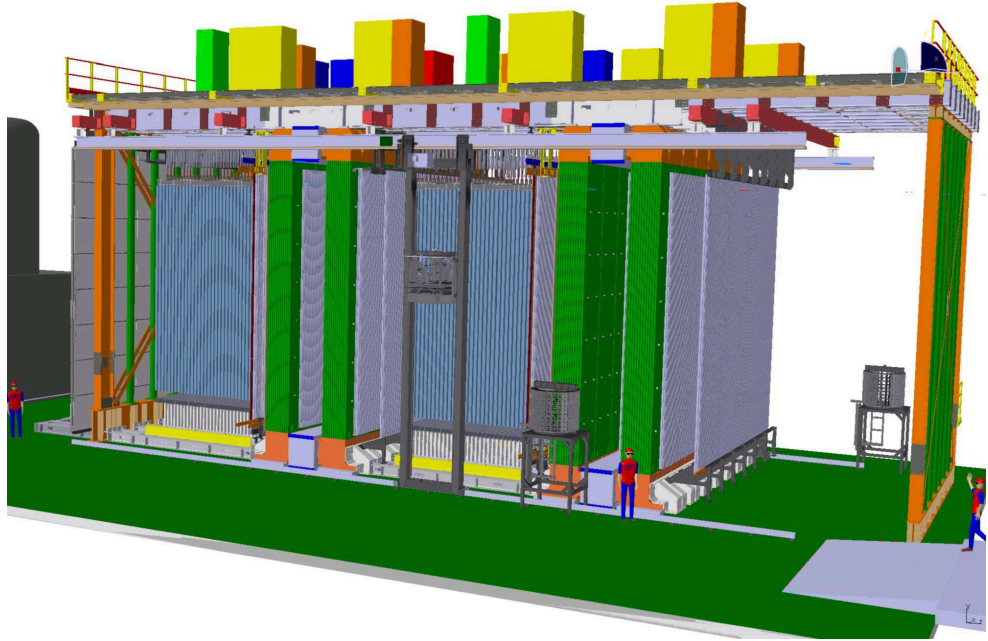}
\end{center}
\caption{\small Schematic drawing of the OPERA detector at LNGS.} 
\label{fig:operadet}
\end{figure}

Each of the two Super Modules  (SM1 and SM2) consists of 103,168 lead/emulsion bricks arranged in 
31 target planes (Fig.~\ref{fig:operadet}), each one followed by two scintillator planes with an effective granularity of 
2.6 $\times$ 2.6 cm$^2$. These planes serve as trigger devices and allow selecting the brick containing a neutrino
interaction. A muon spectrometer at the downstream end of each SM allows to measure the muon charge and momentum. 
A large size anti-coincidence detector
placed in front of SM1 allows to veto (or tag) interactions occurring in the material and in the rock upstream of the target.

The construction of the experiment started in Spring 2003. The first
instrumented magnet was completed in May 2004 together with the first
half of the target support structure. The second magnet
was completed in the beginning of 2005. In Spring 2006 all scintillator planes were installed.
The production of the ECC bricks started in October 2006 with the aim of completing the full target
for the high-intensity run of 2007. 

\begin{figure}[ht]
\begin{center}
\includegraphics[width=8cm]{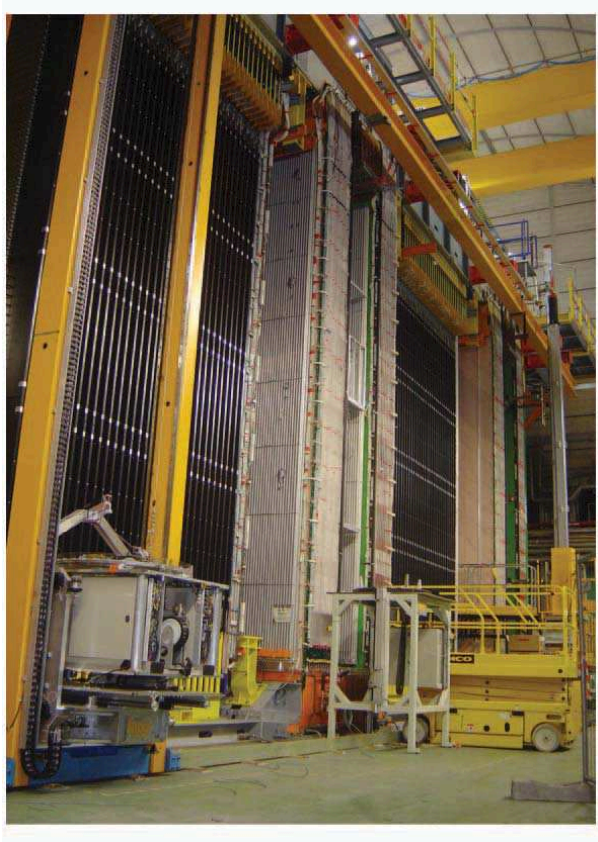}
\end{center}
\vspace{-0.7cm}
\caption{\small Photograph of OPERA in the LNGS Hall C.} 
\label{fig:fig1}
\end{figure}

The run of August 2006 was conducted with electronic detectors only, 
taking neutrino interactions in the rock upstream of the detector, in the passive material of the mechanical 
structure and in the iron of the spectrometers. In addition, the information from a tracking plane made of pairs of emulsion 
films (Changeable Sheets, CS) was used to study the association
between emulsion-film segments with tracks reconstructed in the Target Tracker (TT). 
Fig.~\ref{fig:fig1} shows a photograph of the detector in the underground Hall C of LNGS as it was during the neutrino run.

\subsection{The electronic detectors}

The needs of adequate spatial resolution for high brick finding
efficiency, for good calorimetric measurement of the events,
as well as the requirement of covering large surfaces ($\sim$ 6,000 m$^2$),
impose strong requirements on the TT.  
Therefore, the cost-effective technology of scintillating strips with 
wave length shifting fiber readout was adopted.

The polystyrene scintillator strips are 6.86 m long, 10.6 mm thick and 26.3 mm wide. 
A groove in the center of the strip houses the 1 mm diameter fiber.
Multi anode, 64-pixel photomultipliers are placed at both ends of the fibers. 
A basic unit of the TT called module consists of 64 strips glued together. 
One plane of 4 modules of horizontal strips and one of 4 modules of vertical strips 
form a scintillator wall providing X-Y track information. 
The readout electronics is based on a 32-channel ASIC~\cite{asic} that outputs a charge proportional to the
signal delivered by each pixel of the photomultipliers with a dynamic range from 1 to 100 photoelectrons.

Muon identification and charge measurement are needed for the study of the muonic $\tau$-decay channel and for the suppression 
of the background from the decay of charmed particles, featuring the same topology. 
Each muon spectrometer~\cite{spectro} consists of a dipolar magnet made of two iron arms for a total weight of
990 ton. The measured magnetic field intensity is 1.52 T. 
The two arms are interleaved with vertical, 8 m long drift-tube planes for the precise 
measurement  of the muon-track bending. 
Planes of Resistive Plates Chambers (RPCs) are inserted between the iron plates of the arms, providing a coarse tracking inside 
the magnet, range measurement of the stopping particles and a calorimetric analysis of hadrons.  

In order to measure the muon momenta and determine their sign with high accuracy, 
the Precision Tracker (PT) is built of thin walled aluminum tubes with 38 mm outer diameter and 
8 m length~\cite{hpt}.
Each of the $\sim$ 10,000 tubes has a central sense wire of 45 $\mu$m diameter. 
They can provide a spatial resolution better than 300 $\mu$m. 
Each spectrometer is equipped with six fourfold layers of tubes. 
 
RPCs identify penetrating muons and measure their charge and momentum in an independent way with respect to the PT. 
They consist of electrode plates made of 2 mm thick plastic laminate of high resistivity painted with graphite. 
Induced pulses are collected on two pickup strip planes made of copper 
strips glued on plastic foils placed on each side of the detector. 
The number of individual RPCs is 924 for a total
detector area of 3,080 m$^2$. The total number of digital channels is about 25,000, one for each of 
the 2.6 cm (vertical) and 3.5 cm (horizontal) wide strips.

In order to solve ambiguities in the track spatial-reconstruction each of the two drift-tube planes of the PT upstream of the 
dipole magnet is complemented by an RPC plane with two 42.6$^\circ$ crossed strip-layers called XPCs. 
RPCs and XPCs give a precise timing signal to the PTs. 

Finally, a detector made of glass RPCs 
is placed in front of the first Super Module, acting as a veto system for interactions occurring in the upstream rock. 
The veto detector was not yet operational for the August 2006 run. The PT was in the commissioning phase
with two working planes. 
The TT and the RPCs already passed a full commissioning with cosmic-ray muons before the 
run\footnote{The cosmic muon flux in the LNGS Hall C integrated over the full solid angle
is about 1 muon/m$^2$/hour.}.

OPERA has a low data rate from events due to neutrino interactions well localized 
in time, in correlation with the CNGS beam spill. The synchronization with the spill is done offline via GPS. 
The detector remains sensitive during the inter-spill time and runs in a trigger-less mode. 
Events detected out of the beam spill (cosmic-ray muons, background from environmental radioactivity, 
dark counts) are used for monitoring.
The global DAQ is built as a standard Ethernet network whose 1,147 nodes are 
the Ethernet Controller Mezzanines  plugged on controller boards interfaced 
to each sub-detector specific front-end electronics.  A general 10 ns clock 
synchronized with the local GPS is distributed to all mezzanines in order to insert a
time stamp to each data block. The event building is performed by sorting individual subdetector data by their time stamps.

\subsection{Emulsion films, bricks and related facilities}

An R\&D collaboration between the Fuji Company and the Nagoya group 
allowed the large scale production of the emulsion films needed for the experiment 
(more than 12 million individual films) fulfilling the requirements of uniformity of response and of production, time stability, 
sensitivity, schedule and cost~\cite{emulsions}.  The main peculiarity of the emulsion films used in high energy physics compared to normal photographic films is the relatively large thickness of the sensitive layers ($\sim$ 44 $\mu$m) 
placed on both sides of a 205 $\mu$m thick plastic base.

A target brick consists of 56 lead plates of 1 mm thickness and 57 emulsion films. 
The plate material is a lead alloy with a small calcium content to improve its mechanical properties.
The transverse dimensions of a brick 
are 12.7 $\times$ 10.2 cm$^2$ and the thickness along the beam direction is 7.5 cm (about 10 radiation lengths). 
The bricks are housed in support structures placed between consecutive TT walls. 

In order  to reduce the emulsion scanning load the use of Changeable Sheets, 
successfully applied in the CERN CHORUS experiment~\cite{chorus}, 
was extended to OPERA. CS doublets are attached to the downstream face of each brick and can be 
removed without opening the brick. 
Charged particles from a neutrino interaction in the brick cross the CS and produce a trigger in the TT scintillators. 
Following this trigger the brick is extracted and the CS developed and analyzed in the scanning facility at LNGS. 
The information of the CS is used for a precise prediction of the position of the tracks 
in the most downstream films of the brick, hence guiding the so-called scan-back vertex-finding procedure. 

The hit brick finding is one of the most critical operations for the success of the experiment,
since one aims at high efficiency and purity in detecting 
the brick containing the neutrino interaction vertex. This requires 
the combination of adequate precision of the TT, precise extrapolation and high track finding efficiency in the 
CS scanning procedure. During the neutrino run of August 2006
a successful test of the whole procedure was performed by using an emulsion detector plane consisting 
of a matrix of 15 $\times$ 20 individual CS doublets 
with overall transverse dimensions of 158 $\times$ 256 cm$^2$ inserted in one of the SM2 target planes.

The construction of more than 200,000 bricks for the neutrino target
is accomplished by an automatic machine, the Brick Assembly Machine,
operating underground in order to minimize the number of background tracks from cosmic-rays and environmental radiation. 
Two Brick Manipulating Systems on the lateral sides of the detector position the bricks in the target walls and also
extract those bricks containing neutrino interactions. 

While running the experiment, after the analysis of their CS doublets, bricks with neutrino events
are brought to the LNGS external laboratory, exposed for several hours to cosmic-ray muons for film 
alignment ~\cite{barbuto} and then disassembled. The films are 
developed with an automatic system in parallel processing chains and dispatched to the scanning labs.

The expected number of bricks extracted per running-day with the full target installed and CNGS nominal intensity
is about 30. The large emulsion surface to be scanned requires fast 
automatic microscopes continuously running at a speed of $\sim$ 20 cm$^2$ film surface per hour. 
This requirement has been met after R\&D studies conducted using two different approaches
by some of the European groups of the Collaboration (ESS)~\cite{ees} and by the Japanese groups (S-UTS)~\cite{suts}.


\section {The first run with CNGS neutrinos}
\indent

For a detailed description of the CNGS beam operation during the first run with neutrinos of August 2006 we refer to 
the official CNGS WEB page~\cite{cngs}.
The commissioning of the beam started on 10 July 2006 following a series of technical 
tests of individual components performed from February to May. During this phase
the SPS delivered 7$\times$10$^{15}$ p.o.t., equivalent to 1 hour of CNGS running with nominal
intensity.

\begin{figure}[ht]
\begin{center}
\includegraphics[width=12cm]{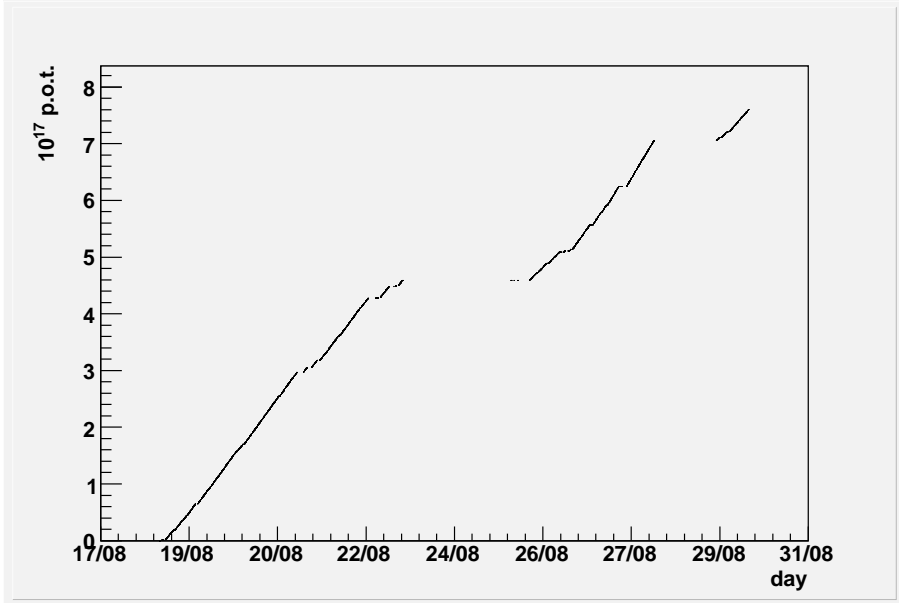}
\end{center}
\vspace{-0.7cm}
\caption{\small The CNGS proton integrated intensity during the August 2006 run.} 
\label{fig:integrated}
\end{figure}

The first shot of the extracted proton beam onto the CNGS target was made on 11 July. A low intensity run with neutrinos
took then place from 18 to 30 August 2006 with a total integrated intensity of 7.6 $\times$ 10$^{17}$ p.o.t.
(Fig.~\ref{fig:integrated}). The beam had been active for a time equivalent to about 5 days. 
The low intensity was partly due
to the chosen SPS cycle and to the intensity of the spill that was 55\% of the nominal value during
the first part of the run and 70\% during the second part.

The GPS clock used to synchronize the CERN accelerators and OPERA
had been fine-tuned before the start of data-taking. 
At CERN the current pulse of the kicker magnet used for the beam 
extraction from the SPS to the TT41 line was time-tagged by a GPS unit with absolute time (UTC) calibration. 
An analogous GPS at the LNGS site provided the UTC timing signal to OPERA.
The resulting accuracy in the time synchronization between CERN and OPERA timing systems was better than 100 ns.
However, during the first days of the run a time offset 
of 100 $\mu$s was observed due to problems in adjusting the time tagging of the kicker pulse. 
This offset was eventually reduced to 600 ns. 

The OPERA detector started collecting neutrino interactions from the very first beam spills 
with nearly all electronic detectors successfully operating. 
Altogether, 319 neutrino events, with an estimated
5\% systematic uncertainty, were taken by OPERA during the August run. 
This is consistent with the 300 expected events for the given integrated intensity of 7.6 $\times$ 10$^{17}$ p.o.t..
The analysis of the CNGS data conducted at CERN  and the comparison with simulations is in progress.
Once completed, we expect to reach a 20\% systematic error on the prediction of the number of muon events from
neutrino interactions in the rock. This error is due to uncertainties in the neutrino flux prediction, in the cross-section
and in the muon transport in the rock.

\begin{figure}[ht]
\begin{center}
\includegraphics[width=14cm]{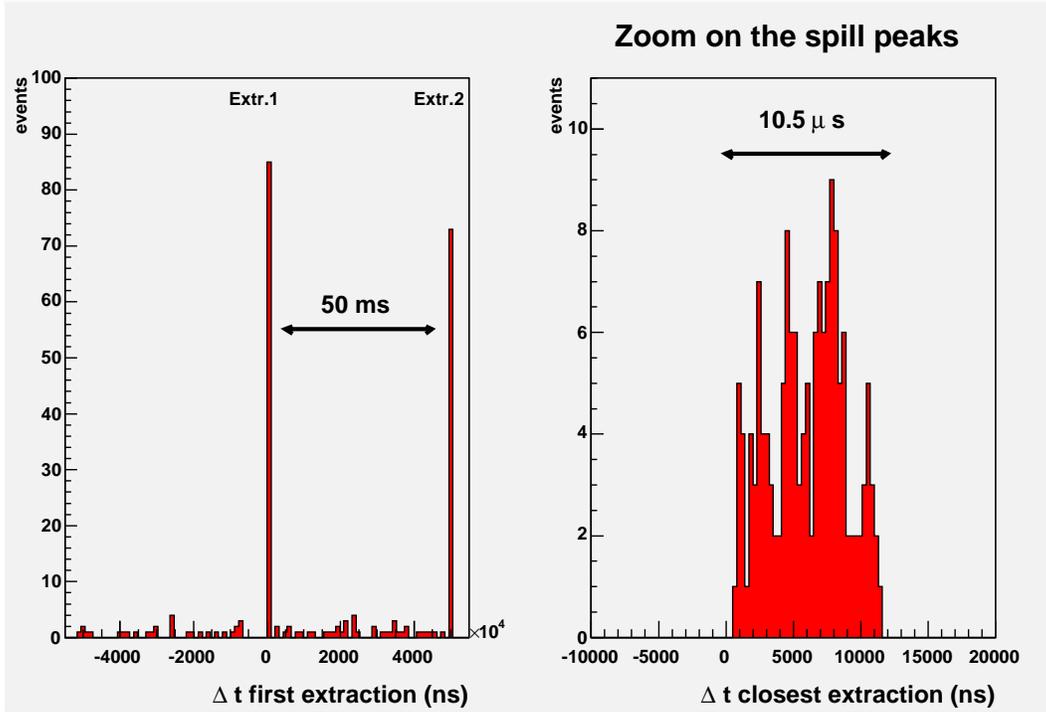}
\end{center}
\vspace{-0.7cm}
\caption{\small Time distribution of events collected in the neutrino run. 
The event time difference with respect to the closest extraction is shown in the right histogram.} 
\label{fig:timebox1}
\end{figure}

\begin{figure}[ht]
\begin{center}
\includegraphics[width=8.5cm]{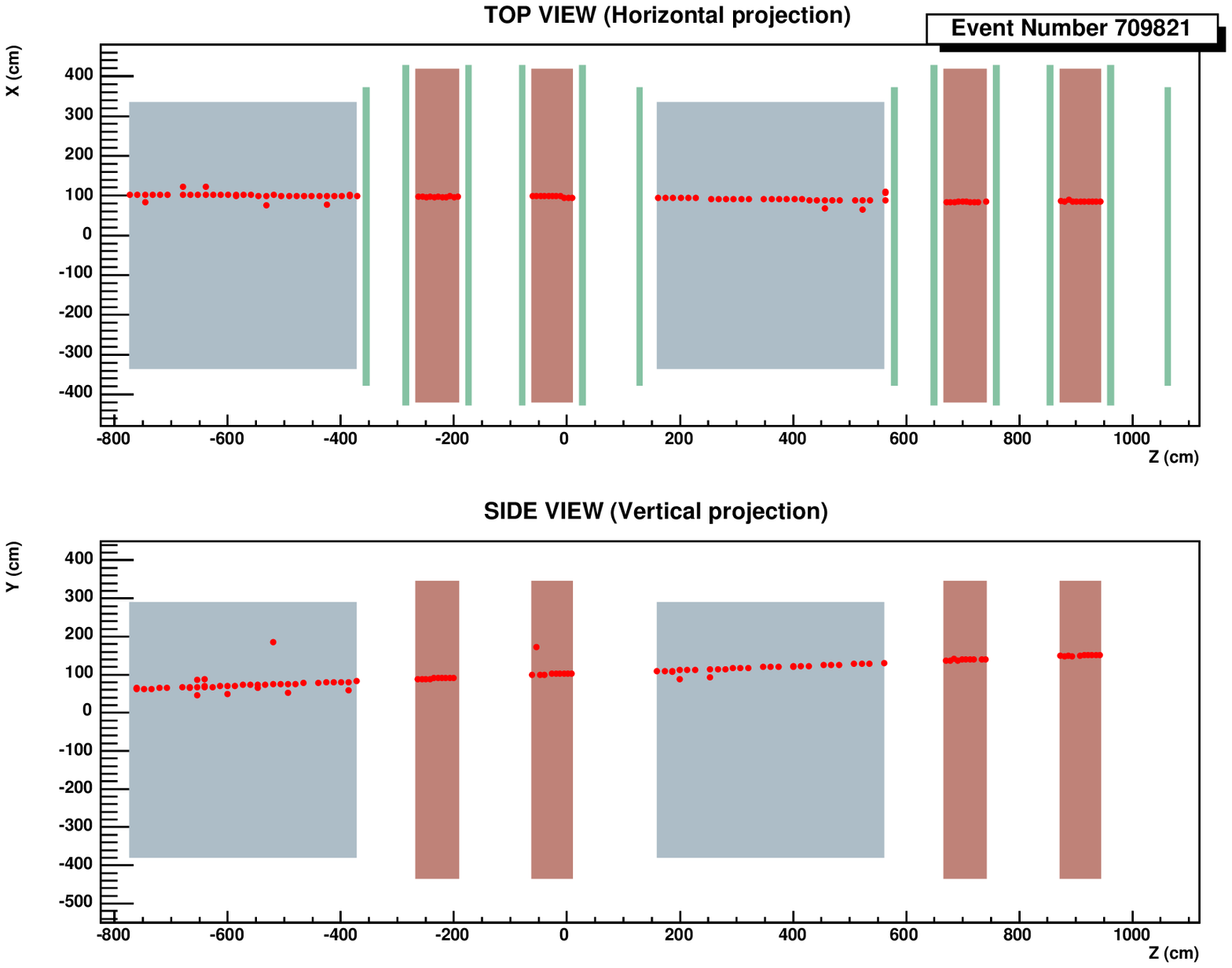}
\includegraphics[width=8.5cm]{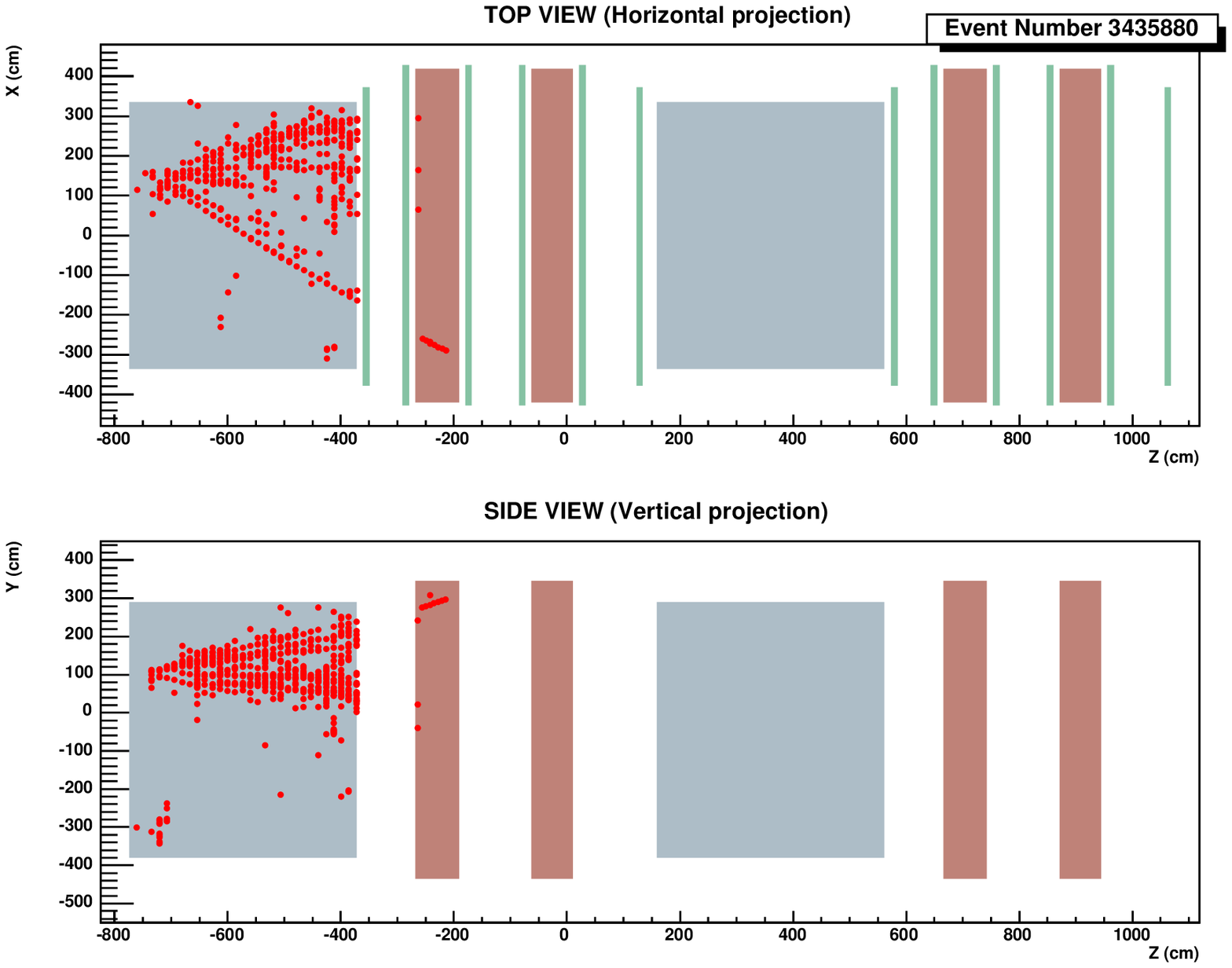}
\includegraphics[width=8.5cm]{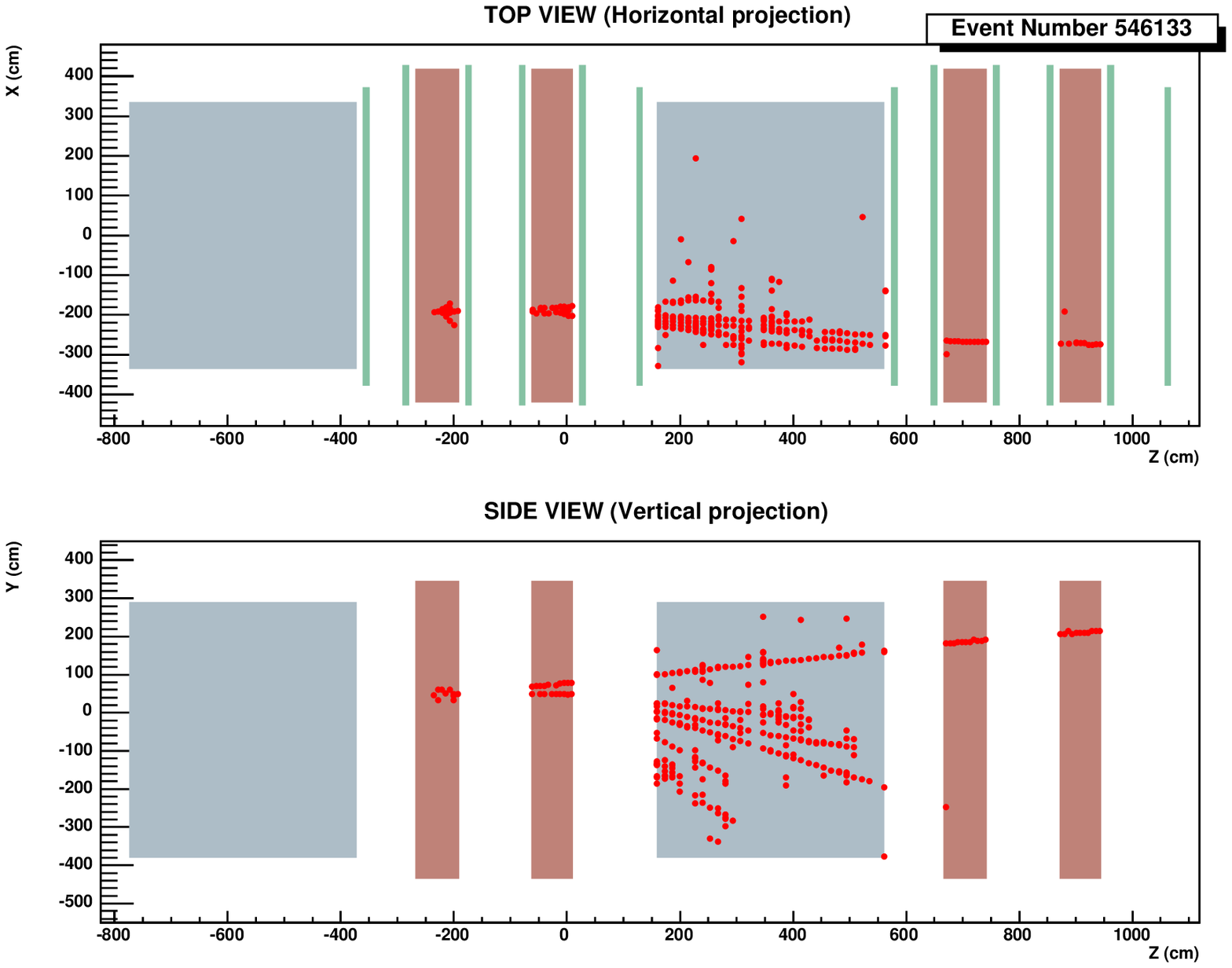}
\includegraphics[width=8.5cm]{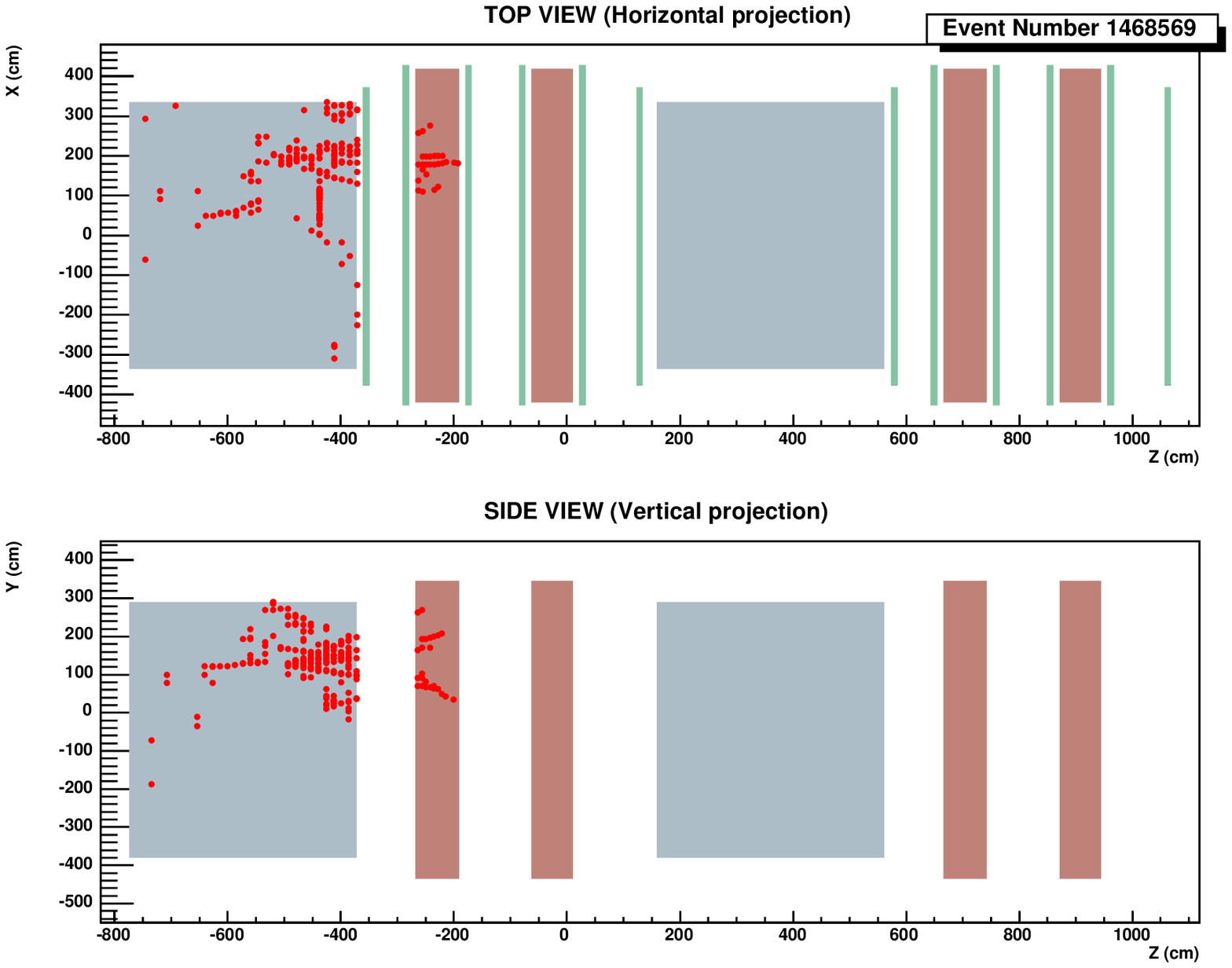}
\end{center}
\vspace{-0.7cm}
\caption{\small Display of neutrino events from the CNGS run. For each event the top and side views are shown, respectively.
The SM targets are indicated in blue, the spectrometers in light brown, TT and RPC hits in red.
See the text for event classification.} 
\label{fig:eventsclean}
\end{figure}

The event analysis was performed in two ways. In the first one the event 
timing information was treated as a basic selection tool, since the time window of beam events
is well sized in a 10.5 $ \mu$s interval, while the uniform cosmic-ray background 
corresponded to 10$^{-4}$ of the collected statistics (Fig.~\ref{fig:timebox1}). 
The second analysis dealt with the reconstruction of track-like events disregarding timing information. 
Neutrino events are classified as: 1) CC neutrino interactions in the rock 
upstream of the detector or in the material present in the hall leading to a penetrating muon track
(Fig.~\ref{fig:eventsclean}, top-left);
2) CC and NC neutrino interactions in the target material (Fig.~\ref{fig:eventsclean}, top-right and bottom-right) 
and CC interactions in the iron of the spectrometers (Fig.~\ref{fig:eventsclean}, bottom-left).
 
\begin{figure}[ht]
\begin{center}
\includegraphics[width=13cm]{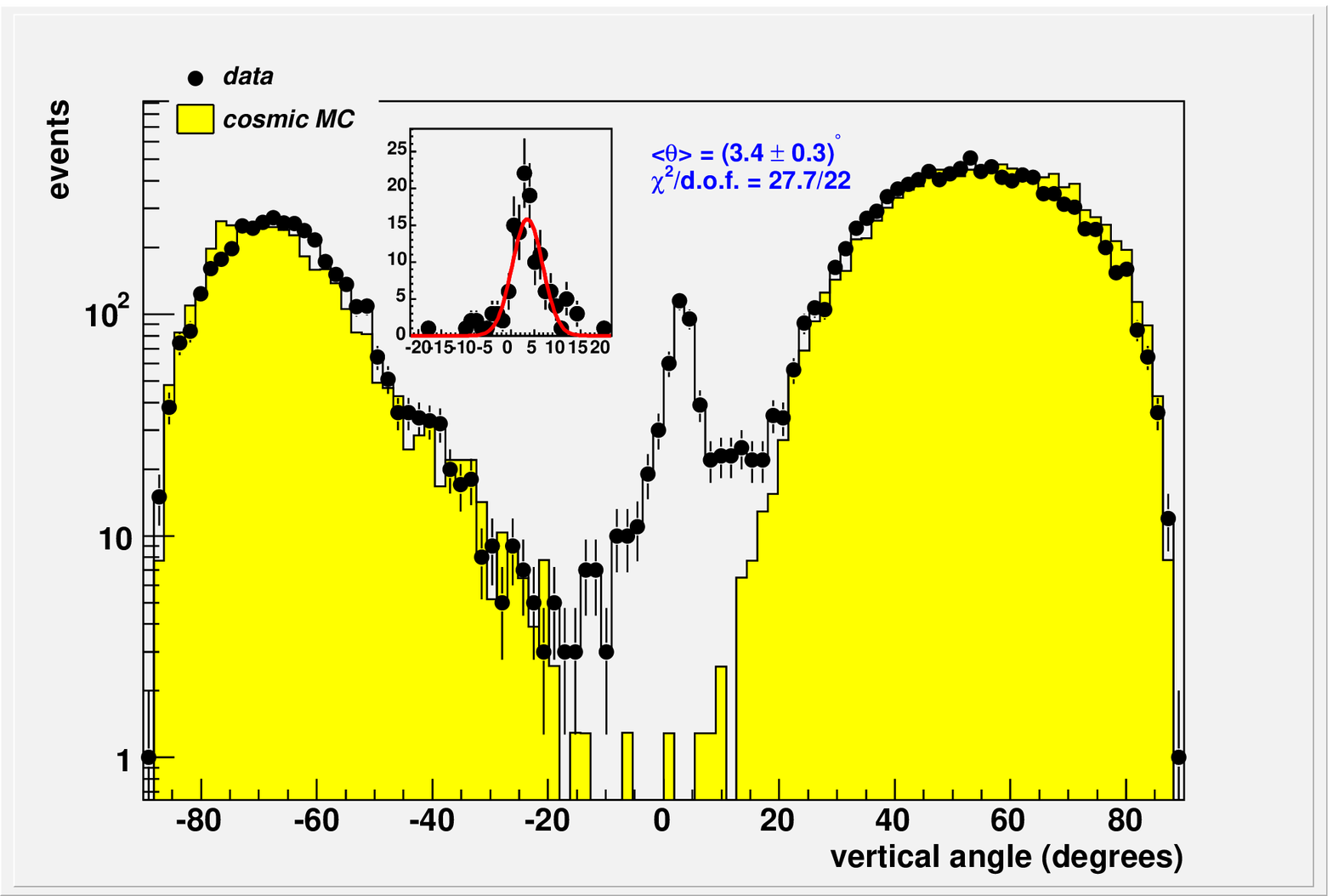}
\end{center}
\vspace{-0.7cm}
\caption{\small Angular distribution of beam-induced and cosmic-muon events taken with the electronic 
detectors (black points).
The histogram indicates the predictions from cosmic-ray simulations~\cite{macro}. 
The inset shows the angular distribution of on-time beam events.} 
\label{fig:cosmics}
\end{figure}

The $\theta$ angular distribution with respect to the horizontal axis obtained by
selecting single-track events is shown in Fig.~\ref{fig:cosmics}. Events were selected with a minimum 
number of 6 layers of fired RPCs in each spectrometer. 
In the same Figure, the distribution of simulated cosmic-ray muons from~\cite{macro} is also shown.   
The comparison between experimental data and Monte Carlo events proved the beam-induced nature
of the muons in the peak around the horizontal direction. 
By counting events selected with topological criteria
we found $\sim$ 10\% of the events corresponding to beam spill data missing in the CERN database.
A Gaussian fit to the $\theta$ angle of the events on-time with the beam (shown in the inset of Fig.~\ref{fig:cosmics}) 
yielded a mean muon angle of 3.4$\pm$0.3$^\circ$ in agreement with the value of  3.3$^\circ$ expected for neutrinos 
originating from CERN and traveling under the earth surface to the LNGS underground halls.
The systematic error on $\theta$ was evaluated to be 3\% by applying different track reconstruction 
models. 
 
\begin{figure}[hb]
\begin{center}
\includegraphics[width=8.6cm]{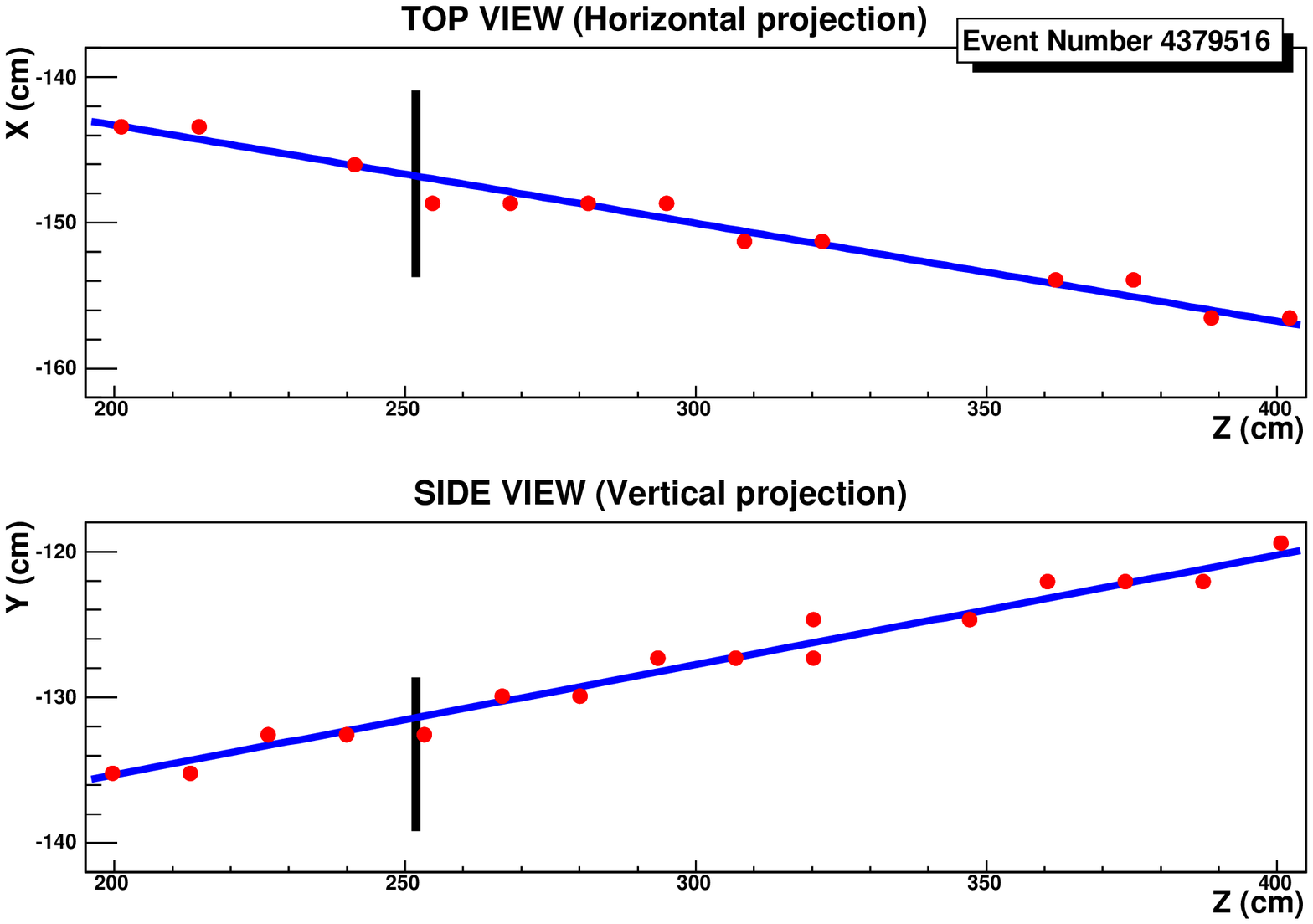}
\includegraphics[width=8.5cm]{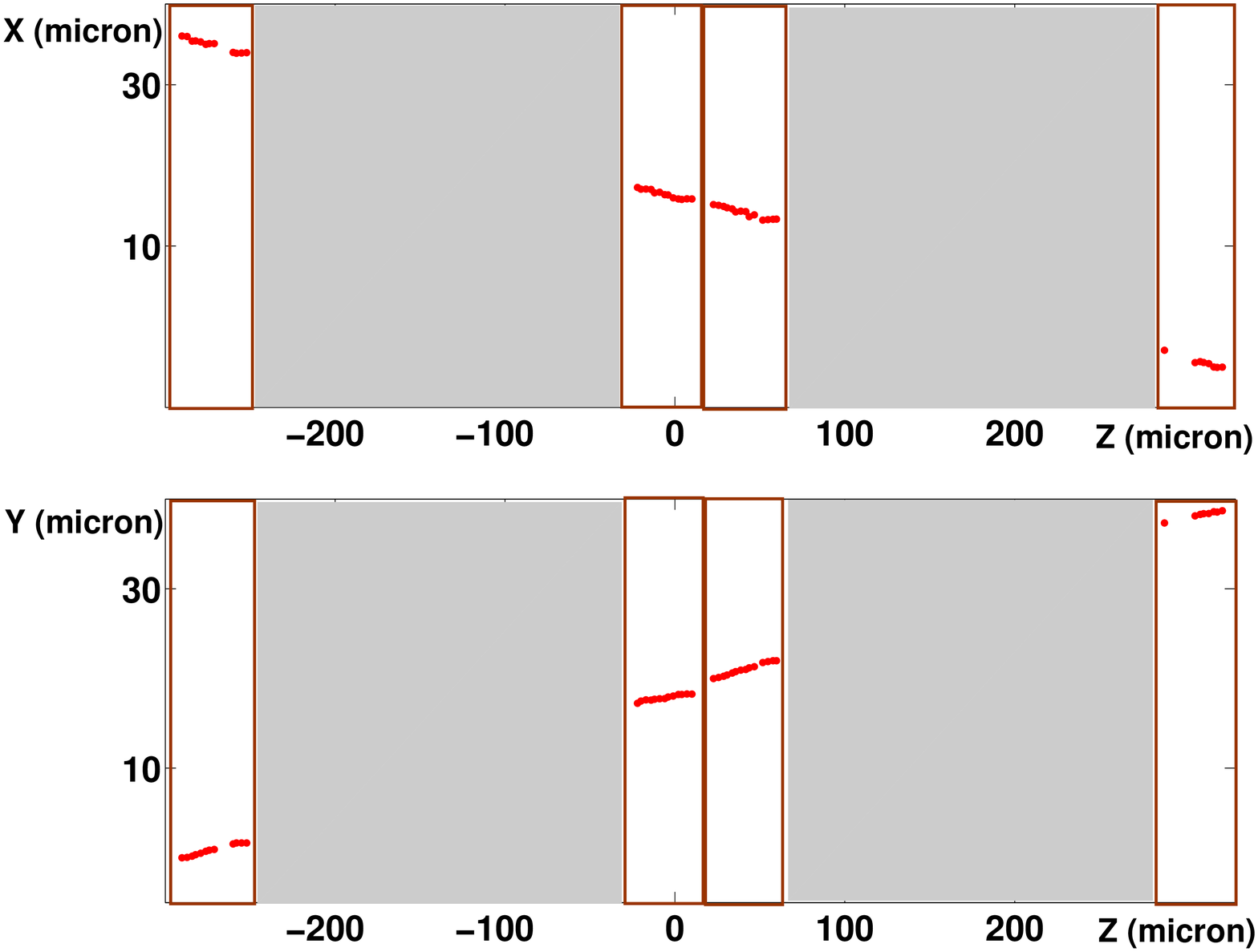}
\end{center}
\vspace{-0.7cm}
\caption{\small Left: display of one event with the muon passing through the CS detector plane.
Only hits of the electronic detectors close to the CS plane are shown;
the vertical segment indicates the position of the CS doublet intercepted by the track. 
Right: display of the corresponding 4 micro-tracks reconstructed in the CS doublet.} 
\label{fig:CSTT}
\end{figure}

Concerning the analysis of the CS detector, 9 muons produced by neutrino interactions in the rock surrounding the 
detector crossed the CS plane surface. 5 muon tracks predicted by the electronic detectors were found by scanning 
the emulsion films. The reasons of inefficiency can be traced-back to the tight cuts applied in this preliminary analysis and 
in the significant decrease of the fiducial volume. In fact,
the dead space between adjacent emulsion films was $\sim$ 10\% and the scanning was only performed 
up to 3 mm from the film edge, bringing the overall dead space to $\sim$ 20\%. 
However, the test proved the capability in passing from the centimeter scale of the
electronic tracker resolution to the micrometric resolution of nuclear
emulsions. The angular difference between predicted and found tracks is
better than 10 mrad, largely dominated by the electronic detector resolution.
Fig.~\ref{fig:CSTT} shows the display of one of the 6 reconstructed events.


\section{Conclusions}
\indent

We reported the first detection of neutrino events from the long baseline CERN CNGS beam with the OPERA experiment 
in the underground Gran Sasso lab. The electronic detectors of the experiment performed successfully 
with an overall data-taking efficiency larger than 95\% during the August 2006 run. 
The scintillator Target Trackers and the
spectrometers equipped with RPCs allowed to identify muon tracks from CC neutrino 
interactions occurring in the rock and in the material upstream of the detector, as well as in the detectors themselves.

319 neutrino-induced events were collected for an integrated intensity of 7.6 $\times$ 10$^{17}$ p.o.t. in agreement with the expectation of 300 events. The reconstructed zenith-angle distribution from penetrating muon tracks 
is centered at 3.4$^\circ$ with a 10\% statistical error, as expected for neutrinos originating from CERN and traveling 
under the earth surface to LNGS.

A test of the association between muon tracks reconstructed with the electronic detectors and with an emulsion detector plane 
was also successfully performed, proving the capability of passing from the centimeter scale of the
electronic tracker resolution to the micrometric resolution of nuclear emulsions. 
The angular difference in the track association is
better than 10 mrad, largely dominated by the electronic detector resolution.

The success of this first OPERA run with CNGS neutrinos is the first step towards the operation of the complete detector. 


\section {Acknowledgements}

We thank CERN for the timely commissioning of the CNGS facility and
for its first successful operation, and INFN for the continuous support given
to the experiment during the construction, installation and commissioning phases through its LNGS laboratory.
We warmly acknowledge funding from our national agencies: 
{\it Fonds National de la Recherche Scientifique et Institut Interuniversitaire 
des Sciences Nucl\'eaires} for Belgium, MoSES for Croatia, IN2P3-CNRS for France, BMBF for Germany, 
INFN for Italy, the {\it Japan Society for the
Promotion of Science} (JSPS), the {\it Ministry of Education, Culture,
Sports, Science and Technology} (MEXT) and the {\it Promotion and Mutual Aid Corporation for Private Schools of Japan} for Japan, SNF and ETHZ for Switzerland. 
We thank INFN for providing fellowships and grants to non Italian researchers.
We are finally indebted to our technical collaborators
for the excellent quality of their work over many years of design, prototyping and construction
of the detector and of its facilities.

\end{document}